\documentclass[%
 reprint,
 amsmath,amssymb,
 aps,
]{revtex4-1}
\usepackage{mathtools}
\usepackage{multirow}
\usepackage[export]{adjustbox}

\usepackage{graphicx}
\usepackage{dcolumn}
\usepackage{bm}
\usepackage{hyperref}
\usepackage[english]{babel}

\usepackage[normalem]{ulem}
\usepackage{comment}
\usepackage[draft]{todonotes}   
\usepackage{xcolor}
\usepackage{subcaption} 
\usepackage{booktabs} 
\usepackage{titlesec}

\titlespacing\section{0pt}{6pt plus 4pt minus 2pt}{4pt plus 2pt minus 2pt}
\titlespacing\subsection{0pt}{12pt plus 4pt minus 2pt}{0pt plus 2pt minus 2pt}
\titlespacing\subsubsection{0pt}{12pt plus 4pt minus 2pt}{0pt plus 2pt minus 2pt}

\newcommand{\ul}[1]{\underline{#1}}
\newcommand{\ub}[1]{\underbar{#1}}
\newcommand{\ubh}[1]{\hat{\underbar{#1}}}
\newcommand{\ubhb}[1]{\hat{\bar{\underbar{#1}}}}
\newcommand{\ens}[2]{\{\ub{#1}_{#2,e}\}_{e=1}^E}
\newcommand{\defeq}{\overset{\text{\tiny def}}{=}}

\DeclareMathOperator*{\argmin}{arg\,min}

\preprint{APS/123-QED}

\begin{document}
\title{Multirotor Ensemble Model Predictive Control I: Simulation Experiments}
\author{Erina Yamaguchi}
\affiliation{Department of Aeronautics and Astronautics}
\author{Sai Ravela}
\affiliation{Earth, Atmospheric and Planetary Sciences\\ \\
Earth Signals and Systems Group\\
Massachusetts Institute of Technology, Cambridge, MA}

\begin{abstract}
    Nonlinear receding horizon model predictive control is a powerful approach to controlling nonlinear dynamical systems. However, typical approaches that use the Jacobian, adjoint, and forward-backward passes may lose fidelity and efficacy for highly nonlinear problems. Here, we develop an Ensemble Model Predictive Control (EMPC) approach wherein the forward model remains fully nonlinear, and an ensemble-represented Gaussian process performs the backward calculations to determine optimal gains for the initial time. EMPC admits black box, possible non-differentiable models, simulations are executable in parallel over long horizons, and control is uncertainty quantifying and applicable to stochastic settings. We construct the EMPC for terminal control and regulation problems and apply it to the control of a quadrotor in a simulated, identical-twin study. Results suggest that the easily implemented approach is promising and amenable to controlling autonomous robotic systems with added state/parameter estimation and parallel computing.
\end{abstract}

\maketitle

\section{Introduction}

There is tremendous interest in Uncrewed Aircraft Systems (UAS) for autonomous observation of the Earth and Environment~\cite{ravela13,Ravela2018}. The proliferation of low-cost UAS often entails using low-cost autopilots and ``cheap" hardware. Despite their prolific use, there is a need for skillful autonomous flight, resilience in stochastic and adversarial environmental conditions, and efficiency for long-duration observations. Unfortunately, the stock autopilots typically contain many tunable parameters in PID controllers, and despite recent efforts at low-cost adaptive control~\cite{goel2021a,lee2021adaptive}, there remains a strong need for efficacious optimal control.

Model predictive control can address some needs~\cite {MAYNE20142967}, especially with the rise of high-performance parallel-distributed embedded computing. In this framework, a model of the dynamical system optimizes control inputs by considering the dynamical system's evolution up to a future horizon. To the degree that the model is skillful about the future, it ameliorates the limitations of classical feedback control, including LQR/LQG approaches for multi-stage closed-loop control, mainly when systems must operate far away from equilibrium. In contemporary practice, the models are linear and identified empirically from data or reduced from governing equations. Highly nonlinear systems might require dense linearization and many forward-backward iterations of the (discrete) Hamilton-Jacobi-Bellman equations that solve the optimal control problem for multi-stage two-point boundary value problems~\cite{BrysonHo69}. These problems' horizons are typically short, placing additional computational demands on control cycle efficacy.

If one were to pursue fully nonlinear receding horizon model predictive control, which could admit longer horizons, then from the many similar inverse problems employing a terminal quadratic performance index and adjoining the dynamical model as a constraint, we know that control inputs computed for every step in the window, but discarding most for application to the plant before repeating the cycle. Additionally, computing adjoints is nontrivial, especially when the forward model is complex and may contain non-differentiable elements. Most nonlinear MPC is, of course, wholly unsuited in stochastic settings.

Can we overcome these limitations? Is there an approach that could handle nonlinear model predictive control, apply to stochastic settings, require no adjoint computations, apply away from equilibrium and offer the stability of LQR/LQG near equilibrium? Remarkably, the answer is yes from developments in ensemble approaches to state and parameter estimation. 

This paper presents a different approach to receding horizon nonlinear model predictive control. An ensemble of nonlinear models with identical parameters and initial conditions is simulated to a finite fixed horizon using initial (prior) control input perturbations at the start (current) time. The performance index yields a Gaussian process as the Normal equation that provides the backward-in-time (adjoint) calculations, setting the optimal gain to update the control ensemble (posterior). The plant receives control inputs selected from the posterior control ensemble, and the process repeats. This way, optimal control closed-loop terminal controllers that bring the system close to
the desired conditions at an unknown terminal time, regulators maintaining the system around a desired
state condition, and trajectory tracking controllers are feasible. With the advent of embedded parallel computing (e.g., GPUs), this approach is feasible for developing highly skilled controllers.

The distinguishing feature of the proposed approach is that the forward-backward iteration is unlike the open-loop iterations of model predictive control or optimization. The ensemble controller admits fully nonlinear, black-box, possibly non-differentiable models and requires no linearization. Using the ensemble to a fixed terminal time, it quantifies and uses uncertainty to produce a statistical adjoint, optimal gain, or transfer function for control. It is amenable, as constructed, for deterministic or stochastic control. Because the EMPC approach leverages and provides state and control uncertainties, it is possible to apply information gain arguments to judiciously select control inputs or models~\cite{trautner2020informative}, which is further valuable for targeting both resource-constrained and redundant-actuation settings.

Autonomous observing systems are the application setting for the present work in ensemble nonlinear model predictive control, which we envision as the core of Lapponica Pilot, our ongoing autopilot development project. In context, the ensemble approach is applicable not just for control but also to continually estimate parameters and states of the nonlinear aeromodel~\cite{stengel,beard,luukkonen2011modelling}, enabling highly skilled real-time adaptive nonlinear model predictive controllers. Furthermore, since the governing model equations will likely remain imperfect, we develop Neural Dynamical Systems~\cite{trautner2020informative,trautner2019neural} where the physical model pairs with a well-sized neural network~\cite{trautner2021learn,liziwei} and the parameters of both are ``trained" online using the ensemble approach~\cite{Ravela_2021b}. Thus, in operation, Lapponica executing on embedded-GPU boards will pair with stock hierarchical PID~\cite{px4,px4reference,px4HolybroPixhawk} until the training yields an independent autopilot.

Here, we take the first step, developing EMPC and verifying feasibility in a simulated identical-twin experiment for a quadrotor. In the identical-twin setup, we assume the model is perfect and apply EMPC to another simulation Results indicate that EMPC is highly effective at setting the gains for terminal control and regulation. 

The remainder of this paper is as follows. Section~\ref{sec:rw} describes related work. Section~\ref{sec:empc} develops the ensemble approach to nonlinear receding horizon model predictive control. Section~\ref{sec:exp} describes simulation experiments with a quadrotor. A discussion follows in Section~\ref{sec:disc}, and Section~\ref{sec:concl} concludes the paper and notes future work. 

\section{Related Work}
\label{sec:rw}

The literature contains numerous control schemes for multi-rotors, including PID~\cite{px4,luukkonen2011modelling}, sliding mode and backstepping control~\cite{tripathi}, adaptive PID approaches for multi-rotors~\cite {spencer2021a,goel2021a}, are also increasing. Linear model predictive approaches~\cite{Islam_2017} and nonlinear approaches~\cite{elhesasynlmpcquad} including particle swarm methods~\cite{merabti15} have been proposed, with further application to trajectory planning/tracking~\cite{sakawa99}. Developing a full-fledged Hamilton-Jacobi-Bellman solution for receding horizon nonlinear model predictive control is rarely possible ~\cite{BrysonHo69}. Still, one must discard most of the inputs over the fixed interval. Alternatives such as sequential action methods~\cite{Tzorakoleftherakis19} are preferable. Our approach is distinct from all these, but we note that, as formulated, it only produces the control input ensemble for the start or current time. 

To our knowledge, Ravela and Banks~\cite{Banks_Ravela_2005} coined and formulated Ensemble Control in the sense of  Ensemble Model Predictive Control used in this paper. However,  developing literature also refers to Ensemble Control~\cite{zlotnik2011synthesis,Qi2012EnsembleCO} as controlling many structurally identical systems with parameter differences. Moreover, this approach has been applied, for example, to distributed/networked linear systems. There is some similarity because the ensemble controller, in our case, controls many model simulations, but there is only one single controlled plant. Our approach draws from the ensemble approach to state and parameter estimation~\cite{AnEnsembleKalmanSmootherforNonlinearDynamics,Ravel2007} as a nonlinear model predictive control mechanism and is, as such, distinct.

The ensemble approach draws from extensive work in ensemble approaches to Bayesian state and parameter estimation~\cite{Evensen2022,AnEnsembleKalmanSmootherforNonlinearDynamics}, where it has shown efficacy for nonlinear, high-dimensional problems, quantifying uncertainty, and leveraging parallelism, with numerous variants and connections to non-Gaussian Bayesian estimation. In particular, we use Ravela and McLaughlin's fast ensemble smoother~\cite{Ravela2007} formulation for control and apply the ensemble Kalman transform filter developed in~\cite{Ravela2010,HUNT2007112}.

\section{Ensemble Model Predictive Control}
\label{sec:empc}
This section describes EMPC, which uses a nonlinear calibrated model to generate control inputs at time step $n=0$ (\emph{wlog}). Performance variable projections of a model ensemble forecast $N$ steps away from a fixed initial state and control input ensemble lead to optimal control solutions for the terminal controller and regulator. The approach admits black-box nonlinear models and requires no linearization, and it quantifies uncertainty to statistically model the backward-in-time process required for nonlinear model predictive control. 

This section begins with specific definitions, then describes the ensemble forecast and statistical adjoint calculations, and synthesizes the receding horizon ensemble model predictive controller and the quadrotor dynamics to which it is applied. 
\subsection{Definitions}
\label{ssec:def}
Continous-time dynamical systems of the form $\dot{x_t}=f(x_t,u_t)$ are typically discretized $t_n = n\delta T$. Thus, consider a deterministic discrete-time nonlinear system with corresponding observation and performance operators:

\begin{eqnarray}
\underbar{x}_{n+1} &=& F(\underbar{x}_{n},\underbar{u}_n;\underline{\alpha}), \\
\underbar{y}_n &=& h(\underbar{x}_n,\underbar{u}_n; \underline{\beta} )+\underline{\nu},\\
\underbar{z}_n & = & g(\underbar{x}_n,\underbar{u}_n;\underline{\gamma})+\underline{\epsilon},
\end{eqnarray}
where, $\underbar{x}_n$ are the state and measurement $\underline{y}_n$ vectors at time step $n$, $F$ is the state equation or forward model with parameter vector $\underline{\alpha}$, and $h$ is the measurement equation with parameter vector $\underline{\beta}$. The measurement noise is assumed here to be time-invariant (homoskedastic) and Gaussian distributed, i.e., $\underline{\nu}\sim \mathcal{N}(\underline{0}, C_{yy})$. We specify the performance variable $\underbar{z}_n$, usually for a terminal time step $n=N$, with mapping $g$ and parameter vector $\underline{\gamma}$, which the model may stochastically satisfy with a precision $\underline{\epsilon} \sim \mathcal{N}(\underbar{0}, C_{zz})$ that is unchanged in any time step. Note that $F$, $h$, and $g$ are deterministic nonlinear functions, but the states, parameters, and control inputs are generally unknown and represented with uncertainty. 

\paragraph{The N-step Ensemble Forecast:} A special case of interest is an $N$-step simulation of the model equations (state function) with a zero-order hold,  holding the control input at step $n=0$ for the $N$ steps. In this scenario, two mappings are useful. The first map is the $N$-step forward propagation:
\begin{eqnarray}
{\ub{x}}_N &\defeq& F_N(\ub{x}_0;\ub{u}_0;\ul{\alpha}),\\
&=&\underbrace{F(\ldots F(F}_{N\;\text{times}}(\ub{x}_0,\ub{u}_0;\ul{\alpha}),\ub{u}_0;\ul{\alpha})\ldots,\ub{u}_0;\ul{\alpha}).
\end{eqnarray}

The second mapping is the projection of the state variable into the performance variable space, written for time-step $N$, as:
\begin{eqnarray}
    \hat{z}_N &\defeq& G_N(\ub{x}_0,\ub{u}_0;\ul{\kappa}),\\
      &=& g(\ub{x}_N, \ub{u}_0;\ul{\gamma}),
\end{eqnarray}
where $\ul{\kappa} = [\ul{\alpha}; \ul{\gamma}]$ (note: we use ``matlab notation" to stack column vectors). 

\paragraph{Sampling Distributions:} In the present ensemble model predictive control approach, a Gaussian distribution $\ub{u}_{0,e} \sim \mathcal{N}(\bar{\ub{u}}_{0}, C_{u_0u_0})$ with a population mean (trim) $\bar{\ub{u}}_{0}$ and covariance $C_{uu}$ generates a control input ensemble $U_0\defeq \ens{u}{0}$ of size $E$ every control cycle and evolves over the control task. 

Representing the ensemble as a matrix $U\equiv U_{D\times E}$ in which each $D$-dimensional ensemble member occupies a column estimates its distributional parameters. In particular, the ensemble mean is $\ubhb{u}_0\defeq \frac{1}{E} \sum_{e=1}^E \ub{u}_{0,e}$. Letting the control perturbation as $\tilde{\ub{u}}_{0,e}\defeq \ub{u}_{0,e}-\ubhb{u}_{0}$ and defining the perturbation ensemble $\tilde{U}_0\defeq \ens{$\tilde{u}$}{0}$ yields the unbiased ensemble estimate of the covariance:
\begin{equation}
    \hat{C}_{u_0u_0} = \frac{1}{E-1} \tilde{U}_0 \tilde{U}^T_0.
\end{equation}

An ensemble forecast up to a horizon $N$ steps away, followed by the projection, is written as 
\begin{equation}
    \hat{Z}_N \defeq G_N(\ub{x}_0, U_0; \ul{\kappa}),
\end{equation}
which is short-hand for simulating $\forall e$, $\hat{z}_{N,e} = G_N(\ub{x}_0,\ub{u}_{0,e})$ and constructing the ensemble $\hat{Z}_N = \ens{$\hat{z}$}{N}$. From the projected ensemble, the sample covariance matrices $\hat{C}_{\hat{z}_N\hat{z}_N}$ and the cross-covariance matrix
\begin{equation}
    \hat{C}_{\hat{z}_Nu_0} = \frac{1}{E-1}\tilde{\hat{Z}}_N\tilde{U}_0^T,
\end{equation}
are constructed. Note that $\hat{C}_{\hat{z}_n\hat{z}_n}\neq C_{zz}$ for any $n$. The former is the covariance of the model ensemble, while the latter is the conditional covariance controlling the precision with which the goals may be satisfied. We assume that in the limit of the large ensemble, all the ensemble estimates converge to the population parameters. 

\subsection{Forward-Backward Gaussian Processes}

Let us consider the Taylor expansion of the simulation and projection operator from an initial condition with a control perturbation, assuming it exists. That is
\begin{eqnarray}
    \ubh{z}_{N} &=& G_N(\ub{x}_0, \bar{\ub{u}}_0+\tilde{\ub{u}}_{0};\ul{\kappa})\\
    &=& G_N(\ub{x}_0, \bar{\ub{u}}_0;\ul{\kappa})+\mathcal{G} \tilde{\ub{u}}_{0}.
\end{eqnarray}

Thus, 
\begin{eqnarray}
    \mathcal{G} \tilde{\ub{u}}_{0} &=& \ubh{z}_{N} - G_N(\ub{x}_0, \bar{\ub{u}}_0;\ul{\kappa})\\
    &=& \tilde{\ubh{z}}_{N} + \bar{\ubh{z}}_N -G_N(\ub{x}_0, \bar{\ub{u}}_0;\ul{\kappa})\\
    &=&\tilde{\ubh{z}}_{N} + \delta\ubh{z}_N.\label{eq:taylor}
\end{eqnarray}
Where, $\bar{\ubh{z}}_N = \left<\ubh{z}_{N}\right>$ is the expectation (ensemble mean) of the nonlinearly propagated control input random variable $\ub{u}_0=\bar{\ub{u}}_0+\tilde{\ub{u}}_{0}$ and $\delta\ubh{z}_N= \bar{\ubh{z}}_N-G_N(\ub{x}_0, \bar{\ub{u}}_0)$. An estimate for $\mathcal{G}$ emerges as:
\begin{eqnarray}
    \mathcal{G}\left<\tilde{\ub{u}}_0\tilde{\ub{u}}_0^T\right> &=& \left<(\tilde{\ubh{z}}_{N} + \delta\ubh{z}_N)\tilde{\ub{u}}_0^T\right>\\
    &=&\left<\tilde{\ubh{z}}_{N} \tilde{\ub{u}}_0^T\right>,
\end{eqnarray}
as the performance variable bias is uncorrelated with the control input 
perturbation statistics. The expectations are just the associated covariances. Thus, the forward Gaussian process $\mathcal{G}$ is:
\begin{eqnarray}
     \mathcal{G}   &=& C_{\hat{z}_Nu_0}C_{u_0u_0}^{-1}.
     \label{eq:linmod}
\end{eqnarray}
It is important to note that $C_{\hat{z}_Nu_0}$ involves nonlinear ensemble forecast and projection. A backward Gaussian process is also similarly defined:
\begin{eqnarray}
     \mathcal{A}   &=& C_{u_0\hat{z}_N}C_{\hat{z}_N\hat{z}_N}^{-1}.
     \label{eq:backol}
\end{eqnarray}
Thus, 
\begin{equation}
    C_{\hat{z}_N\hat{z}_N} = \mathcal{G} C_{u_0u_0} \mathcal{G}^T. 
\end{equation}

In practice, we apply the ensemble approximation to the forward and backward Gaussian processes. Thus, Equation~\ref{eq:linmod} has an ensemble approximation: 
\begin{equation}
     \hat{\mathcal{G}}   = \hat{C}_{\hat{z}_Nu_0}\hat{C}_{u_0u_0}^{-1},
\end{equation}    
which will be central to the development of the ensemble controller in its square-root form.

\subsection{Ensemble Control}

Consider the performance objective
\begin{eqnarray}
    J(\ub{u}_0|\ub{z}_N) &=& \frac{1}{2}(\ub{z}_N - \ubh{z}_N)^T C_{zz}^{-1} (\ub{z}_N - \ubh{z}_N) \nonumber\\
&&+ \frac{1}{2}(\ub{u}_0 - \bar{\ub{u}}_0)^T C_{u_0u_0}^{-1} (\ub{u}_0 - \bar{\ub{u}}_0),
\end{eqnarray}
where $\ubh{z}_N = G_N(\ub{x}_0, \ub{u}_0;\ul{\kappa})$. 
We interpret the performance objective from a Bayesian standpoint, where the first term on the right-hand side is the likelihood of the controller meeting the performance criterion, and the second term is a prior on the control input at the initial time. Seeking a Maximum a-Posteriori (MAP) solution yields the cost function. This treatment for Gaussian random variables is optimal~\cite{kay1993fundamentals}, but the minimum variance and well-chosen cost function justifications are also common. Here, the Bayesian form is essential for transitioning from deterministic optimal control formulations to the stochastic version and ensemble formulation. When solved, the cost function updates the nominal trim setting $\bar{\ub{u}}_0$. We solve the cost function by minimizing $J$:
\begin{eqnarray}
    \ub{u}^+_0 &=& \argmin_{\ub{u}} J(\ub{u} |\ub{z}_N)
\end{eqnarray}
\paragraph{Nonlinear Controller:} The well-known solution~\cite{Evensen2022,Ravela2007,bouquet11,AnEnsembleKalmanSmootherforNonlinearDynamics,HUNT2007112} can be written as follows: 
\begin{eqnarray}
    \ub{u}^+_0 &=& \bar{\ub{u}}_0 + C_{u_0u_0}\mathcal{G}^T (\mathcal{G} C_{u_0u_0} \mathcal{G}^T + C_{zz})^{-1} (\ub{z}_N - \hat{\ub{z}}_N)\nonumber\\
    &=& \bar{\ub{u}}_0 + C_{u_0\hat{z}_N} (C_{\hat{z}_N\hat{z}_N} + C_{zz})^{-1} (\ub{z}_N - \hat{\ub{z}}_N)\label{eq:meaneq}\\
&=& \bar{\ub{u}}_0 + \mathcal{K}(\ub{z}_N - \hat{\ub{z}}_N)
\end{eqnarray}    
To be sure, Equation~\ref{eq:meaneq} expresses the update to the mean control input at time step $0$ due to a terminal performance criterion at a time step $N$. It requires nonlinear model prediction and projection but uses a linear backpropagation of error through optimal gain $\mathcal{K}$. It is thus fully nonlinear in $\ub{u}_0$ and requires no linearization. As has been shown~\cite{Evensen2022}, it is applicable in the ensemble setting to each ensemble member, i.e.,
\begin{eqnarray}
    \ub{u}^+_{0,e} &=& {\ub{u}}_{0,e} + \hat{C}_{u_0\hat{z}_N} (\hat{C}_{\hat{z}_N\hat{z}_N} + C_{zz})^{-1} (\ub{z}_{N,e} - \hat{\ub{z}}_{N,e}),\label{eq:enseq}\\
    &=&{\ub{u}}_{0,e} + \hat{\mathcal{K}}(\ub{x}_0,\ub{U}_0) (\ub{z}_{N,e} - \hat{\ub{z}}_{N,e})\\
&&\forall e \in \{1\ldots E\}\nonumber 
\end{eqnarray}

Equation~\ref{eq:enseq} is nonlinear, substituting the population cross-covariances with the ensemble estimates. Akin to perturbed observations~\cite{bouquet11,HUNT2007112}, the variable $\ub{z}_{N,e}$ allows for using perturbed performance variables. Still, formulations without perturbations are also feasible, and one can set $\ub{z}_{N,e}=\ub{z}_N$. 

\paragraph{Receding Horizon Ensemble Model Predictive Control:} Equation~\ref{eq:enseq} can be written in matrix form~\cite{Ravela2007,Evensen2022}, producing the N-step ensemble model predictive controller. 
\begin{eqnarray}
    {U}^+_0 &=& U_0 + \tilde{U}_0 \tilde{\hat{Z}}_N^T (C_{\hat{z}_N\hat{z}_N} + C_{zz})^{-1} (Z_N - \hat{Z}_N)\\
       &=& U_0 + \tilde{U}_0 \hat{\mathcal{Z}}_N\\
       &=& U_0 (I_{E\times E} + \hat{\mathcal{Z}}_N)\\
       &=& U_0 \mathcal{M}_N.\label{eq:enscontroller}
\end{eqnarray}

The optimal ensemble controller (Equation~\ref{eq:enscontroller} applies directly in the MIMO setting. Thus, using an initial control input ensemble $U_0$, our approach propagates the model states to horizon $N$, collects performance errors, and produces a message $\mathcal{M}_N$ of size $E\times E$ to produce the optimal posterior ensemble ${U}^+_0$. The plant receives some function of the posterior control ensemble, and the process continues as the horizon recedes. 

\paragraph{Control Selection} ${U}^+_0$ estimates the posterior mean $\hat{\bar{\ub{u}}}^+_0$ and covariance $\hat{C}^+_{u_0u_0}$. Ideally, the mean $\hat{\bar{\ub{u}}}^+_0$ is applied to the plant, but particular ensemble members or other ordered statistics might be preferable in certain situations. For example, the median is a robust estimate when outliers enter the ensemble, often as an artifact of poor sample size. Another extension is to save the entire numerical simulation over each time step up to the horizon, then use the state vector at the time with the best performance from the ensemble member to construct the ensemble forecast. Such a scheme is useful to promote minimum-time solutions. 
 \begin{figure*}[htb!]
    \centering
    \includegraphics[width=7in]{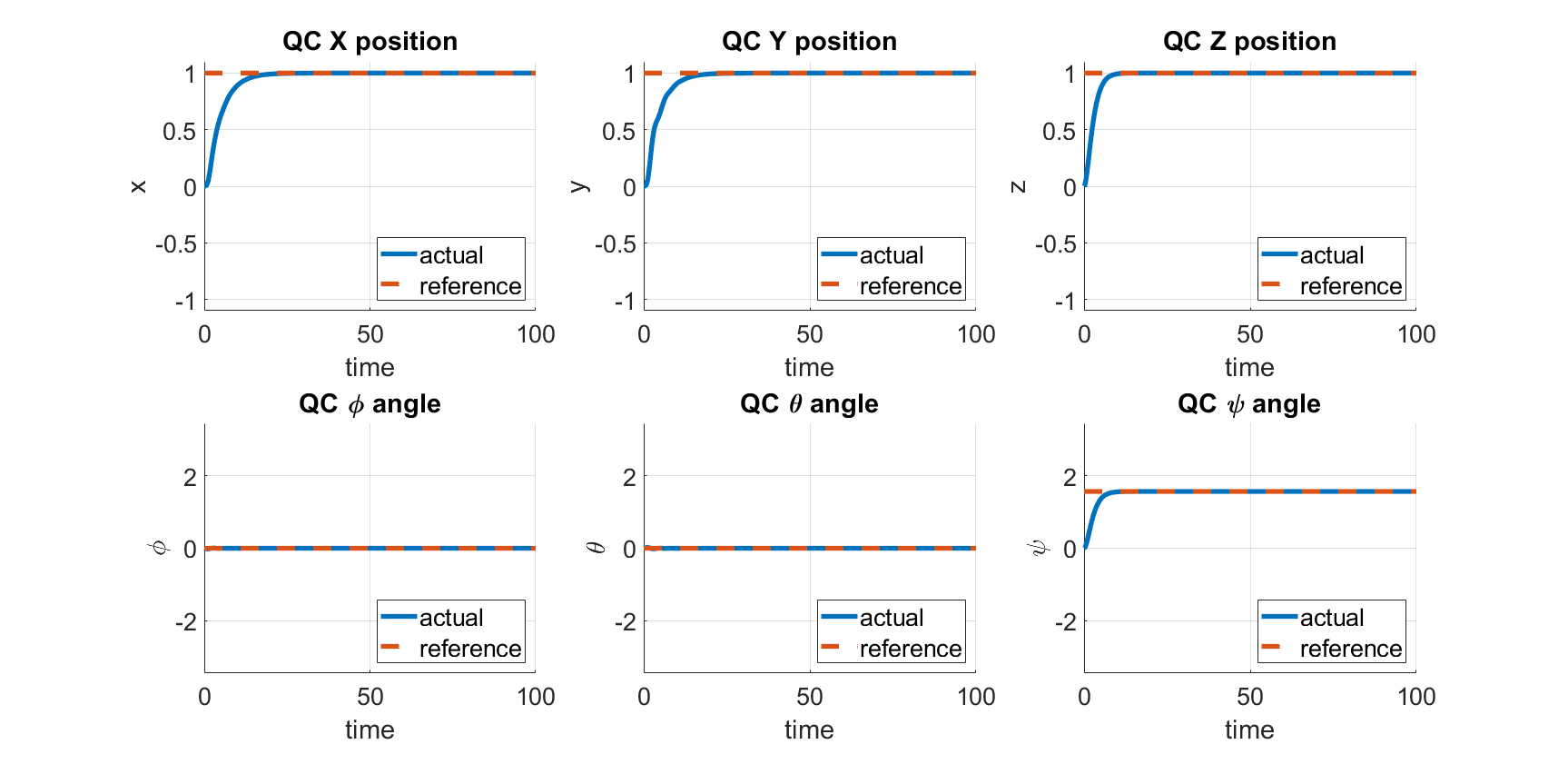}
    \caption{Some state variables as a function of time in the ($x=0,y=0,z=0$)$\Rightarrow$ ($1,1,1$) and ($\phi=0,\theta=0,\psi=0$)$\Rightarrow$ ($0,0,\pi/2$) joint position-yaw fixed-point terminal control problem. The dashed red line marks the reference values in the panel. The state variable trajectory (blue) executes a critically damped motion with optimally and automatically determined ensemble gain. Y-axis units are length(m) and time(s).}
    \label{fig:Exp1State}
\end{figure*}
\paragraph{Gaussian Process Adjoint:} The ensemble controller uses a Gaussian process approximation to the adjoint. It is exact if the dynamics and performance operators are linear, the uncertainties remain Gaussian, and we consider the infinite limit of the ensemble size. In the nonlinear setting, Equation~\ref{eq:enscontroller}, the covariance ($C_{\hat{z}_N\hat{z}_N}$) and cross-covariance $(C_{u_0\hat{z}_N})$ are nonlinear functions of the initial condition and control inputs and hence define a Gaussian process regression. The Gaussian process approximations in the forward direction ($\mathcal{G}$) and the backward direction ($\mathcal{A}$) are approximations to the forward and backward Kolmogorov processes associated with the nonlinear dynamical system $G_N$.

\paragraph{Square-root Forms:} The $\mathcal{M}_N$ calculation often proceeds in square-root form and does not require explicitly constructing the covariance matrices, which can become quite large for certain systems. For example, suppose that $C_{zz} = \rho^2 I_{O\times O}$ for an $O$-sized performance vector variable. Further, suppose that $\tilde{\hat{Z}}_N = \mathbf{U}\mathbf{\Sigma}\mathbf{V}^T$ is the singular value decomposition. Then, we may express Equation~\ref{eq:enscontroller} as~\cite{HUNT2007112,Ravela2010}
\begin{eqnarray}
 {U}^+_0 &=& U_0 + \tilde{U}_0 \tilde{\hat{Z}}_N^T (C_{\hat{z}_N\hat{z}_N} + C_{zz})^{-1} (Z_N - \hat{Z}_N)\nonumber \\
&=&    U_0 + \tilde{U}_0 \mathbf{V}\mathbf{\Sigma}\mathbf{U}^T (\mathbf{U}\mathbf{\Sigma}^2\mathbf{U}^T + \rho^2 I)^{-1} \delta Z_N\\
&=&U_0 + \tilde{U}_0 \mathbf{V}\mathbf{\Sigma}(\mathbf{\Sigma}^2 + \rho^2 I)^{-1}\mathbf{U}^T \delta Z_N\\
&=& U_0 \underbrace{\left[I_{E\times E} +\mathbf{V}\mathbf{\Sigma}(\mathbf{\Sigma}^2 + \rho^2 I)^{-1}\mathbf{U}^T \delta Z_N \right]}_{\mathcal{M}_N}\label{eq:ensvd}
\end{eqnarray}

Equation~\ref{eq:ensvd} is a simplified version where the inversion of the diagonal matrix $(\mathbf{\Sigma}^2 + \rho^2 I)$ is simple. We anticipate $\tilde{\hat{Z}}_N$ of size $(D\times E)$ to be reduced rank in specific high-dimensional applications. One may truncate the eigenspectrum to filter a few modes from the ensemble controller $\mathcal{M}_N$.

 \paragraph{Finite-Interval Trajectory Tracking Controller:} Our formulation extends to trajectory tracking controllers with minor modifications akin to Ravela and McLaughlin's forward-backward algorithm~\cite{Ravela2007}. Consider the interval $(0,N]$, within which performance variable $\ub{z}_n$, $0<n\leq N$  is specified is an indicator variable $o_n = 1$ and perform a single ensemble simulation using a state $\ub{x}_0$ and initial control ensemble $U_{0}$, returning $\hat{Z}_1, \hat{Z}_2 \ldots \hat{Z}_N$. Define the control ensemble update as
\begin{equation}
\mathcal{M}^{sel}_n=\left\{    \begin{matrix}
        \mathcal{M}_n & \text{if } o_n = 1,\\
        I_{E\times E} & otherwise
    \end{matrix} \right.
\end{equation}
The trajectory tracking controller$\mathcal{M}^{traj}$~\cite{Ravela2007} is a fixed-interval update with the performance index specified all along the trajectory:
\begin{eqnarray}
    U_0^{traj} &=& U_0\prod_{n=1}^N \mathcal{M}^{sel}_n,\\
    &=&U_0\mathcal{M}^{traj}.
    \label{eq:traj}
\end{eqnarray}

As the system marches in time, constant-time algorithms to recursively modify the ensemble trajectory controller are also feasible, following Ravela and McLaughlin's fixed-lag algorithm for estimation~\cite{Ravela2007}. This paper does not implement trajectory tracking.

\subsection{Quadrotor Dynamics and Control}
\label{sec:quadyn}

The quadrotor state consists of $\ub{x} = [x,y,z,\phi,\theta,\psi, \dot{x},\dot{y},\dot{z},\dot{\phi},\dot{\theta},\dot{\psi}]$. The 
$\xi\defeq (x, y, z)$)-axis defines the quadcopter's linear position. The inertial frame defines the angular positions with Euler angles ($\phi, \theta, \psi$), respectively roll, pitch, and yaw, by the vector $\eta$. The vector $V_B$ represents linear velocities, and the angular velocities by vector $\nu$. The vector $q$ contains the linear and angular position vectors ($\xi$ and $\eta$).

We follow Luukkonen~\cite{luukkonen2011modelling} exactly and state the equations he has derived in the rest of this section to represent the dynamics used. They are readily available in Matlab. lThe Lagrangian $\mathcal{L}$ is the sum of the translational ($E_{trans}$) and rotational ($E_{rot}$) energies minus the potential energy ($E_{pot}$). 
\begin{equation}
\mathcal{L}(q,\dot{q}) = E_{trans}+E_{rot}-E_{pot} 
\end{equation}
\begin{equation}
     =(m/2)\dot{\xi}^T\dot{\xi}+(1/2)\nu^TI\nu-mgz
\end{equation}
The Euler-Lagrange equations output the linear and angular forces of the rotors, relating to the total thrust and torques of the rotors, respectively. First, the linear external forces are
\begin{equation}
    \begin{bmatrix}
        f \\ \tau
    \end{bmatrix}
    = \frac{d}{dt}(\frac{d\mathcal{L}}{{d\dot{q}}})-\frac{d\mathcal{L}}{dq}
\end{equation}
\begin{equation}
    f = m\Ddot{\xi}+mg
    \begin{bmatrix}
        0 \\ 0 \\ 1
    \end{bmatrix}
\end{equation}
The Jacobian matrix J($\eta$) from $\nu$ to $\dot{\eta}$ is
\begin{equation}
    \begin{split}
     & J(\eta) = \\
    & \begin{bmatrix}
        I_{xx} && 0 && -I_{xx}s_\theta \\
        0 && I_{yy}c^2_{\phi}+I_{zz}s^2_\phi && (I_{yy}-I_{zz})c_{\phi}s_{\phi}c_{\theta}\\
        -I_{xx}s_{\theta} && (I_{yy}-I_{zz})c_{\phi}s_{\phi}c_{\theta} && I_{xx}s^2_{\theta}+I_{yy}s^2_{\phi}c^2_{\theta}+I_{zz}c^2_{\phi}c^2_{\theta}
    \end{bmatrix}
    \end{split}
\end{equation}
The rotational energy $E_{rot}$ can be expressed in the inertial frame as
\begin{equation}
    E_{rot} = (1/2)\nu^TI\nu = (1/2)\Ddot{\eta}^TJ)\Ddot{\eta}
\end{equation}
The angular torques are
\begin{equation}
\label{eq:torque}
    \begin{split}
    \tau & = \tau_B = J\Ddot{\eta}+\frac{d}{dt}(J)\dot{\eta}-\frac{1}{2}\frac{d}{d\eta}(\dot{\eta}^TJ\dot{\eta})\\
    & = J\Ddot{\eta}+ C(\eta, \dot{\eta})\dot{\eta}
    \end{split}
\end{equation}

where $C(\eta, \dot{\eta})$ is the Coriolis term. The matrix has the form, 
\begin{equation*}
    C(\eta, \dot{\eta}) = 
    \begin{bmatrix}
    C_{11} & C_{12} & C_{13} \\
    C_{21} & C_{22} & C_{23} \\
    C_{31} & C_{32} & C_{33} \\
    \end{bmatrix},
\end{equation*}
\begin{equation}
    \begin{split}
        C_{11} & = 0\\ 
        C_{12} & = (I_{yy}-I_{zz})(\dot{\theta}c_{\phi}s_{\phi}+\dot{\psi}s^2_{\phi}c_{\theta})\\
        & +(I_{zz}-I_{yy})
        \dot{\psi}c^2_{\phi}c_{\theta}-I_{xx}\dot{\psi}c_{\theta}\\ 
        C_{13} & = (I_{zz}-I_{yy})\dot{\psi}c_{\phi}s_{\phi}c^2_{\theta}\\
        C_{21} & = (I_{zz}-I_{yy})(\dot{\theta}c_{\phi}s_{\phi}+\dot{\psi}s_{\phi}c_{\theta})\\
        & (I_{yy}-I_{zz}) \dot{\psi}c^2_{\phi}c_{\theta}+I_{xx}\dot{\psi}c_{\theta}\\
        C_{22} & = (I_{zz}-I_{yy})\dot{\phi}c_{\phi}s_{\phi}\\ 
        C_{23} & = -I_{xx}\dot{\psi}s_{\theta}c_{\theta}+I_{yy}\dot{\psi}s^2_{\phi}s_{\theta}c_{\theta}+I_{zz}\dot{\psi}c^2_{\phi}s_{\theta}c_{\theta}\\
        C_{31} & = (I_{yy}-I_{zz})\dot{\psi}c^2_{\theta}s_{\phi}c_{\phi}-I_{xx}\Dot{\theta}c_{\theta}\\ 
        C_{32} & = (I_{zz}-I_{yy})(\dot{\theta}c_{\phi}s_{\phi}s_{\theta}+\dot{\phi}s^2_{\phi}c_{\theta})+(I_{yy}-I_{zz})\dot{\phi}c^2_{\phi}c_{\theta}\\
        & +I_{xx}\dot{\psi}s_{\theta}c_{\theta}-I_{yy}\dot{\psi}s^2_{\phi}s_{\theta}c_{\theta} -I_{zz}\dot{\psi}c^2_{\phi}s_{\theta}c_{\theta}\\
        C_{33} & = (I_{yy}-I_{zz})\dot{\phi}c_{\phi}s_{\phi}c^2_{\theta} - I_{yy}\dot{\theta}s^2_{\phi}s_{\theta}c_{\theta}\\
        & -I_{zz}\dot{\theta}c^2_{\phi}s_{\theta}c_{\theta} +I_{xx}\dot{\theta}c_{\theta}s_{\theta}
    \end{split}
\end{equation}

Equation \ref{eq:torque} leads to the differential equations for the angular accelerations, which are 
\begin{equation}
    \ddot{\eta} = J^{-1}(\tau-C(\eta, \dot{\eta})\dot{\eta})
\end{equation}

\section{Simulation Experiments}
\label{sec:exp}

The simulation setup for the quadrotor system uses Matlab's nonlinear model predictive control example to synthesize the state function (nonlinear quadrotor dynamics )that a numerical Runge-Kutta scheme solves the equations discussed in Section~\ref{sec:quadyn}. The Ensemble controller does not neet the Jacobian for nonlinear model predictive control and, instead, uses the ensemble for a Gaussian Process approximation of the adjoint. For this experiment, the parameters of the quadrotor are as follows:

\begin{eqnarray}
    I& =\begin{bmatrix}
       1.2 & 0 & 0\\
       0 &  1.2 & 0 \\
       0 & 0 & 2.3
    \end{bmatrix} kg.m^2&\;\;\text{Inertia}\\
    k & =  1&\;\;\text{lift coeff.}\;\;\\
    l & =  0.25 m&\;\;\text{rotor arm} \\
    m & =  2 kg&\;\;\text{mass}\\
    b & =  0.2&\;\;\text{drag coeff.}\\
    g & =  9.81m/s^2&\;\;\text{gravity}
    \end{eqnarray}
    
\begin{figure}[htb!]
    \centering
    \includegraphics[width=3.6in]{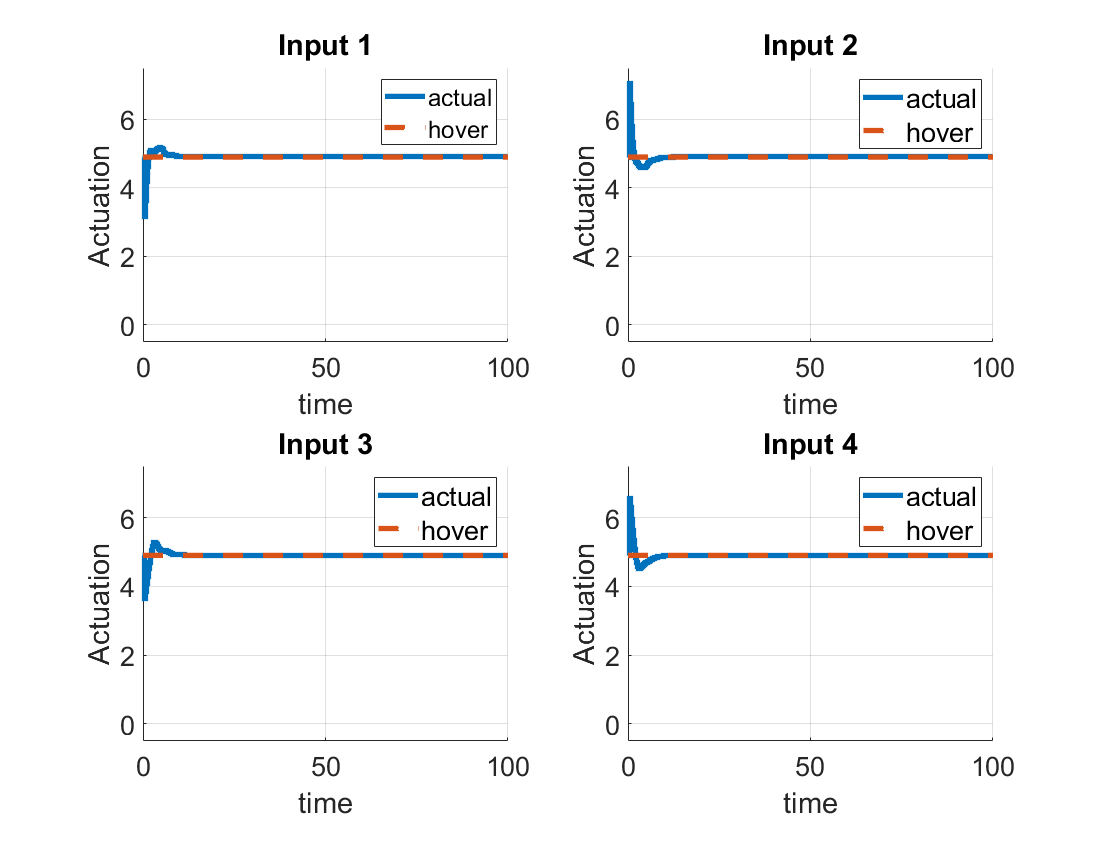}
    \caption{\it The control inputs for each of the four rotors (shown in blue) depart from hover values ($4.9, 4.9, 4.9, 4.9$) (shown in red) on the demand to move to a new location. Due to the horizon's N-step lead, the controller quickly relaxes back to hover following an initial thrust adjustment. The optimal feedback gain is determined automatically from the ensemble model prediction covariance and the performance index without linearizing the forward model or requiring adjoint calculations.}
    \label{fig:Exp1Control}
\end{figure}
We conduct two identical-twin experiments. In an identical twin experiment, one copy is the model reference, and another becomes the ``true" system. The identical-twin setup assumes that the model's initial condition and parameters are identical and share the plant's dynamical equations. We further believe that all state variables are observable. In reality,  the model structure and parameters are generally imperfect. However, we need not know the model (or system) equations; nonlinear black-box models work perfectly well. See Section~\ref{sec:discuss} for additional discussion. With these assumptions, we conduct two ensemble model predictive control experiments. In the first case, we ask the quadrotor to climb, move, and rotate from an initial hovering condition. In the second case, we ask the quadrotor to execute a flight plan that involves visiting a set of waypoints. 

\subsection{Climb, Move, Rotate, and Hover}

\begin{figure}[htb!]
    \centering
    \includegraphics[width=3.5in]{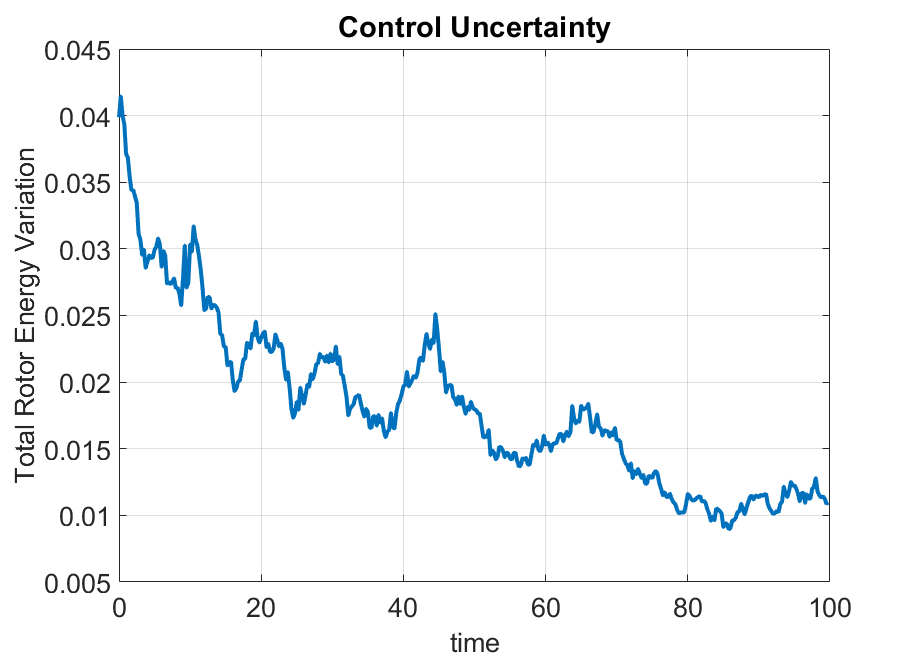}
    \caption{\it The control ensemble's total variance is the sum of rotor control input variances and measures uncertainty in the effort needed to reach the target. This figure plots the square root of the total control input variance as a function of time. It generally decreases, applying only minor perturbations around a trim value near equilibrium, which the performance index's precision partly controls. Ensemble control can quantify the uncertainties (or confidence) in the effort.  }
    \label{fig:varplot}
\end{figure}
The quadrotor, which is in a hover state at position ($0,0,0$) (see Figure~\ref{fig:Exp1State}), is commanded to climb and move to position $(1,1,1)$ and yaw-rotate by $90^o$. The trim setting is the input squared-angular velocities of the rotors at hover $(4.9, 4.9, 4.9,4.9)$ and $E=100$ first-guess perturbations were initially produced {it i.i.d} from a zero mean and standard deviation $0.01$ Normal distribution, which is many more than the needed perturbations. Usually, $16-32$ suffice. From the identical initial condition $\underline{x}_0=(0,0,0,0,0,0,0,0,0,0,0,0)^T$, For this experiment, the time step for numerical simulation is set to $t_n = n\delta T= 0.25n$ (nominally set to $0.1$ in Matlab's state function synthesis). Please note that the Runge-Kutta method uses its time-stepping arrangement internally, but we only access the solution at the end of $\delta T$. We set the horizon to four times the number, i.e., $4\delta T_n  = 1s$. A fourth-order Runge-Kutta method ({\it ode45}) performs the numerical simulations in parallel. The predicted state is the performance variable. Setpoint terminal control requires the quadrotor steadily hover at its new position and orientation. The performance variables have error tolerances arranged in a column vector as $C_{zz} = \ul{\rho}^2 I_{12\times 12}$, where $\rho = 0.001$.

  The ensemble control law~\eqref{eq:enscontroller} uses the $N-step$ prediction and terminal error to update the first-guess control input ensemble. Our control selection scheme uses the median control input (for robustness) to apply to the identical-twin quadrotor simulation proxy of the plant. The simulation then advances a time step. The sampling distribution from the posterior control ensemble variances generates the first-guess control input perturbations for the next control cycle. This way, a 4-step horizon recurs over time, providing control inputs at the beginning of the window to apply to the plant. At no point are the model equations linearized or any adjoint calculations directly performed. 

\begin{figure*}[htb!]
    \centering
    \includegraphics[width=6.5in]{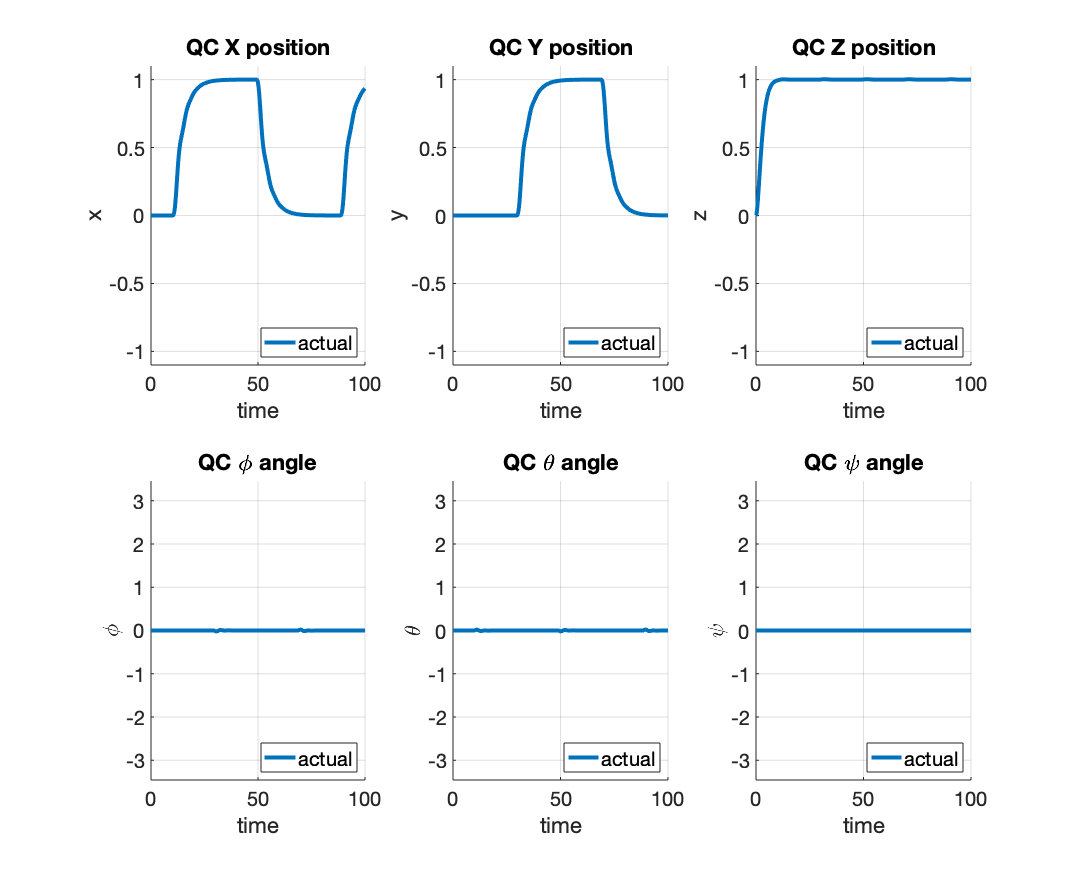}
    \caption{\it Position and attitude variable time-series for Waypoint following sequence around a square $(0,0,1)$, $(1,0,1)$, $(1,1,1)$, and $(0,1,1)$. The reference values are $1$ or $0$ for horizontal positions, $1$ for vertical positions, and $0$ for all attitude angles and rate variables.}
    \label{fig:exp2state}
\end{figure*}

Phenomenologically, from the initial condition, the ensemble of state predictions accelerates control input toward the goal. The state forecast ensemble arrives at the destination much sooner than the plant, suitably decelerating the control inputs. The approach works well even in the presence of bias between the goal and the forecast (state vector) mean. As previously noted, we can modify the minimum energy terminal control formulation with the N-step look-ahead to minimum time by changing the control selection criteria to be applied to the plant and resampling control inputs at the next iteration.

Figure~\ref{fig:Exp1State} shows the position and attitude time series converging relative to the desired goal. Please note that no gain settings exist in the feedback to tune. The propagates error backward, and this is an approximation to the adjoint using a Gaussian process, albeit in reduced-rank square-root form. Figure~\ref{fig:Exp1Control} shows that the ensemble controller initially applies a sharp kick to the rotors. Then the thrust reduces quickly to the background hover values as it nears the goal without oscillation, which is the expected and desired behavior. Figure~\ref{fig:varplot} shows the square root of the predicted control ensemble variance $(rad/s)^2$. Notice that the initial standard deviation of $0.01 \times 4$  generally drops, settling at approximately $0.004$ per rotor. That does not occur immediately upon the quadrotor reaching the goal by $15s$ or so but on a slower timescale. Nonetheless, we are more confident that the rotors are in a steadier rotation regime over time.  

\subsection{Waypoint Following}
In the second experiment, the aircraft climbs and flies through four-way points in a closed square pattern for the available simulation time. It starts at position $(0,0,0)$ and travels to waypoints $(0,0,1)$, $(1,0,1)$, $(1,1,1)$ and $(0,1,1)$ before returning to the first waypoint and repeating. The attitude variables are required to be $0$. In this experiment, each position, attitude variable, and their rates in the state vector members have identical initial tolerance $0.001$. The simulation time step is set to $\delta T = 0.125s$ (half the previous experiment), and the horizon is eight timesteps (one second long). Otherwise, the remaining setup is identical. 
\begin{figure}
    \centering
    \includegraphics[width=3.5in]{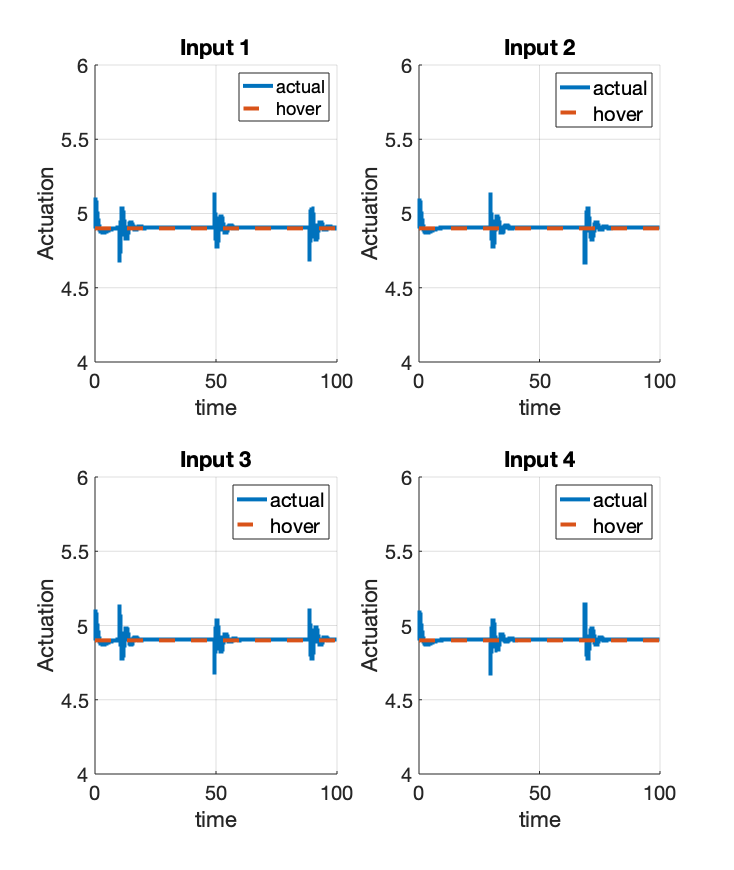}
    \caption{\it The control inputs for each of the four rotors (shown in blue) depart from hover values ($4.9, 4.9, 4.9, 4.9$) (shown in red) on demand to execute a waypoint-following flight-plan. The optimal feedback gain is determined automatically from the ensemble without linearizing the forward model or requiring adjoint calculations. Control inputs cycle quickly to thrust or turn the quadrotor.}
    \label{fig:exp2control}
\end{figure}

Figure~\ref{fig:exp2state} shows the time series of the position and attitude variables. In each case, there are no gains to tune; the ensemble sets the optimal feedback gain at the set horizon. Figure~\ref{fig:exp2control} shows the corresponding rotor commands issued. Again, they are suitably compact in time, as needed for quick trajectory changes.

\begin{figure*}
    \centering
    \includegraphics[width=3.5in]{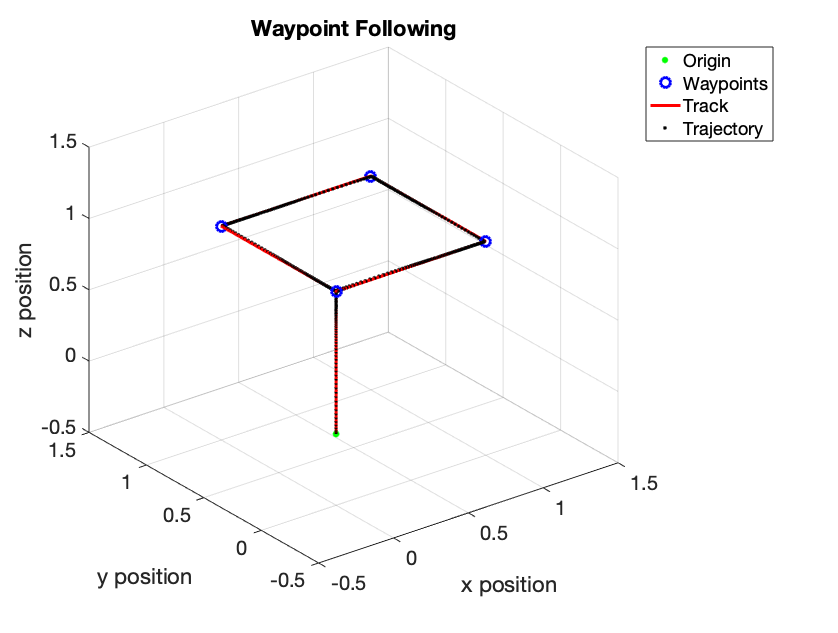}
      \includegraphics[width=3.5in]{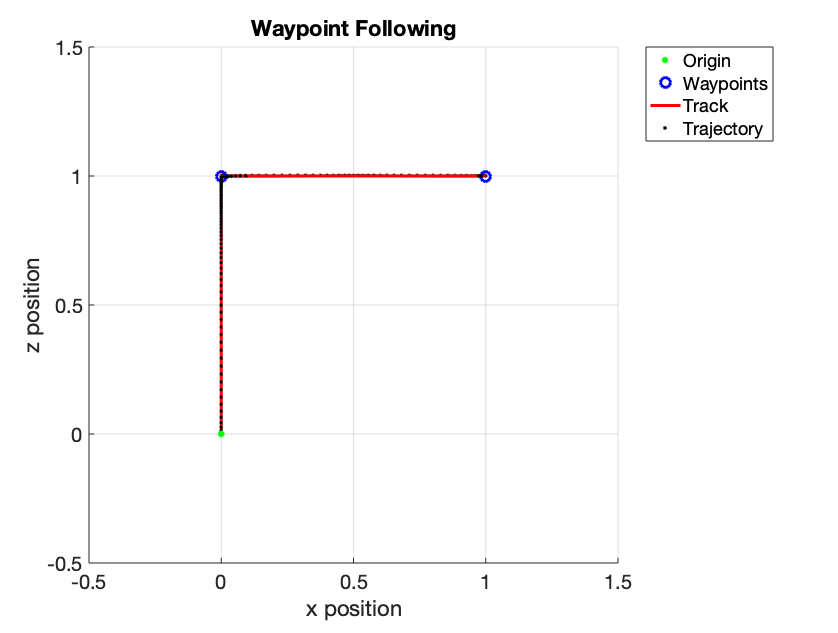}
        \includegraphics[width=3.5in]{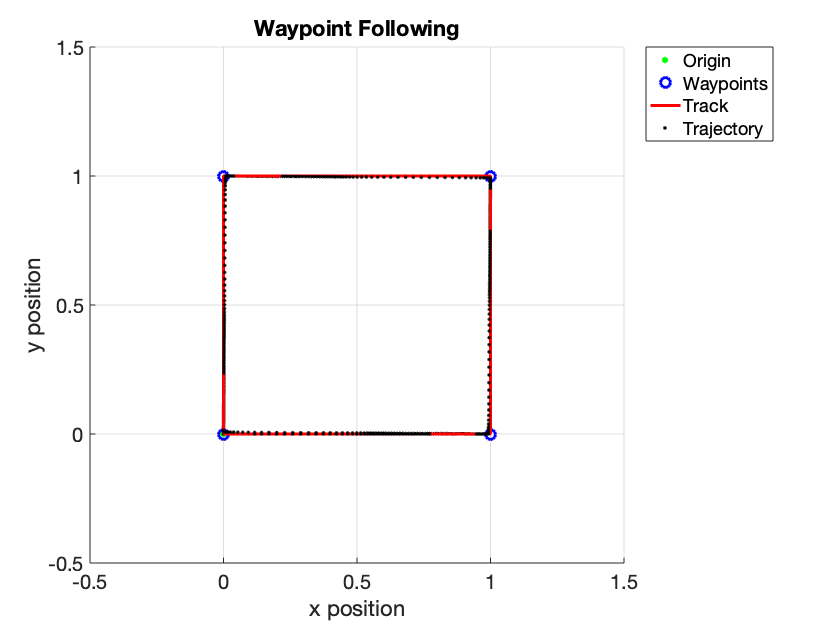}
        \includegraphics[width=3.5in]{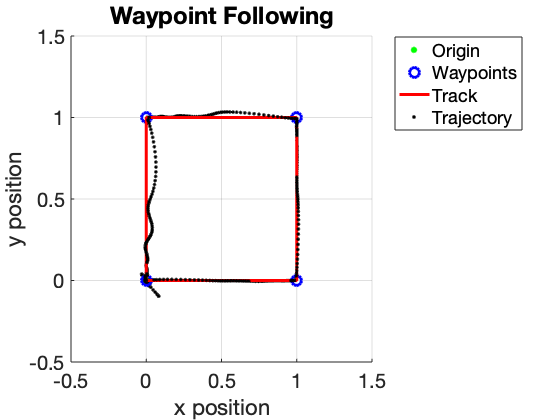}
    \caption{\it Waypoint following. The top left figure shows the waypoints (blue) from the initial location (green) and track (red) to ideally follow. The trajectory flown (black dots) show excellent Waypoint following (and not trajectory tracking). The next two panels show the same information in 2D projections. The last panel shows that a weak performance criterion can degrade the flight path. See text.}
    \label{fig:exp2track}
\end{figure*}

Defining a {\it track} as the straight line path between two waypoints (shown in red in Figure~\ref{fig:exp2track}) and the trajectory flown as dotted black curves, we notice from different axes projections of the flight path that the ensemble controller follows the track quite well. The present ensemble control formulation is for fixed-point terminal control or regulation. The quadrotor flies to a waypoint and switches to the next when it satisfies a goal condition (the Mean Absolute Error (MAE) should be less than $0.001$ over the performance variables). The quadrotor is not trajectory tracking but generally flies straight-line paths and, with a well-selected performance index, performs repeatably well on this task. However, this need not be the case. As the bottom right panel of Figure~\ref{fig:exp2track} shows, when we relax the error tolerances for $\ul{\rho}[1:2] = 0.015$, the trajectories can slip, as shown, depicting departures and oscillations in both terminal and regulation modes. An unstable coupling between the control input ensemble and performance index precision can develop.  

\section{Discussion}
\label{sec:disc}
As systems become increasingly nonlinear and high-dimensional, approaches that rely on linearization, adjoints, and forward-backward passes become inefficient. In many cases, the nonlinearities may not be differentiable. Hamilton-Jacobi-Bellman equations of two-point boundary value problems from which most receding horizon model predictive control follow in exact or approximate form~\cite{BrysonHo69} also face great difficulty incorporating uncertainties in stochastic settings, as noted in optimization, inverse problems, and state and parameter estimation problems~\cite{AnEnsembleKalmanSmootherforNonlinearDynamics,Evensen2022}.

The Ensemble controller estimates a Gaussian process for the computation of backward-in-time (equivalently, adjoint). It quantifies and exploits uncertainty for optimal control, i.e., the gain for closed-loop nonlinear model-predictive control. The approach requires no linearization, which permits longer horizons and admits black-box non-differentiable models. 

While the square-root formulations lead to efficient implementations in one sense, efficiency concerns remain in simulating an ensemble. However, since ensemble simulations are ``embarrassingly parallelizable," the advent of embedded parallel computing may enable real-time applications. GPUs, for example, have shown orders of magnitude improvements for ensemble simulations~\cite{Niemeyer_2014}, and using low-power embedded GPU architectures such as Tegra or Xavier or mixed GPU-FPGA architectures would enable the real-time application. EMPC could be viable for nonlinear model predictive control.

Ensemble sampling requires some care because the ensemble determines $\hat{\mathcal{G}}$  and $\hat{\mathcal{K}}$, which means that a poor choice will lead to the catastrophic failure of the controller. Many choices in the literature, for example, related to the unscented Kalman filter~\cite{unscentedkf} or sigma point methods~\cite{sigmapoint}, are effective at sampling the ensemble. Larger values of $\rho$, as Figure~\ref{fig:exp2track} shows, can lead to poor paths to the terminal, though an explicit path constraint is not present in terminal control or regulation. One must exert additional care to prevent an ensemble collapse~\cite{bouquet11}, which occurs in specific regulator problems where a single performance variable value can artificially drive the ensemble to numerical instability. Perturbing the performance variable in a single update or over time is one way to prevent this. 

The perfect Model assumption could be more problematic. However, we note that nonlinear aero-models~\cite{stengel,luukkonen2011modelling,beard} are likely more skillful than LTI or LTV approximations and more representative of the coupled dynamics than hierarchical PID control. The ensemble approach is just as applicable to recursive state and parameter estimation~\cite{Evensen2022}, so a common computational core can simultaneously provide state and parameter estimates with ensemble control for adaptive control. Further, if physically-based models pair with data-driven models, e.g., neural networks, then the joint estimation of hybrid physical-neural models~\cite{Ravela_2021b} can help address structural deficiencies in the nonlinear model while finding efficacy executing on tensor/GPU architectures~\cite{Niemeyer_2014}.

We note that the ensemble approach informs solutions to many inverse problems in optimization, estimation, control, or learning. For example, the ensemble approach is practical for informative deep learning~\cite{trautner2020informative}. Thus, 
control input uncertainties can form an informative basis for control selection. Even more interesting is that because it quantifies uncertainty, schemes to select the control inputs in resource-constrained or redundant settings optimally become feasible using notions of information gain~\cite{trautner2020informative,raveladddas}.

\section{Conclusions and Future Work}
\label{sec:concl}

An ensemble approach to nonlinear receding horizon model predictive control offers several advantages over classical nonlinear receding horizon model predictive control. They include the admitting black-box models with no linearization. The ensemble approach provides a Gaussian process approximation to the nonlinear model to calculate gains. This paper addresses terminal control and regulation problems with a proposed extension to trajectory tracking. Quadrotor simulation experiments demonstrate excellent performance with automatically determined optimal gains.

We will further develop trajectory control, implement the controller in the PX4 environment for flight tests, and advance the computational core for state and parameter estimation and nonlinear model predictive control.

\section*{Data Availability}
Coded examples related to this paper are available from https://github.com/sairavela/EnsembleControl.git

\section*{Acknowledgment}
Corresponding author: Sai Ravela (ravela@mit.edu). The authors acknowledge support from ONR (N00014-19-1-2273), ARA (S-D00243-05-IDIQ-MIT under ARFL  FA9453-21-9-0054), Liberty Mutual (029024-00020), and the MIT Weather Extreme and CREWSNET Climate Grand Challenge projects. The authors thank Thelonious Cooper for supporting implementation and Prof. Dennis Bernstein for encouraging EMPC.

\bibliographystyle{abbrvnat}
\bibliography{refs}

\end{document}